\documentclass[aip]{revtex4-1}

\usepackage{bm}
\usepackage{amsmath}

\begin{document}

\title{Monte Carlo implementation of \\ a guiding-center Fokker-Planck kinetic equation} 

\author{E. Hirvijoki$^{1}$, A. Brizard$^{2}$, A. Snicker$^{1}$, and T. Kurki-Suonio$^{1}$}
\affiliation{$^{1}$Department of Applied Physics, Aalto University, FI-00076 AALTO, FINLAND \\ $^{2}$Department of Physics, Saint Michael's College, Colchester, VT 05439, USA}


\date{\today}

\begin{abstract}
A Monte Carlo method for the collisional guiding-center Fokker-Planck kinetic equation is derived to include the effects of background magnetic-field nonuniformity. It is shown that, in the limit of a homogeneous magnetic field, the guiding-center Monte Carlo collision operators reduce to the standard Coulomb operators. The coefficients required for the implementation of the method are also given. 
\end{abstract}


\maketitle 

\section{Introduction}
\label{sec:intro}

Understanding and analyzing the behavior of minority energetic particles in the complex magnetic field of a tokamak or stellarator plasma often requires accurate orbit-following simulations. To develop numerically attractive methods, Lie-transformations were first adopted to eliminate fast gyromotion from single particle Hamilton equations~\cite{littlejohn82} and, later, successfully applied to eliminate the gyromotion from the collisional kinetic equation and Fokker-Planck collision term~\cite{brizard2004}. Orbit-following implementations (e.g., ASCOT~\cite{ascot3} and OFMC~\cite{ofmc}) apply the results of Lie-transformations using the guiding-center equations of motion, but traditionally treat the collisional part of the kinetic equation with Monte Carlo operators based on particle phase-space~\cite{boozer1981}.

Several approaches to construct gyrokinetic collision operator have been reported but the Monte Carlo implementations of these operators tend to treat only the velocity part of the phase-space~\cite{xu1991,abel2008,tessarotto1994} or, if the Monte Carlo operator for the spatial diffusion is considered~\cite{vernay}, the collisional part of the kinetic equation is treated separately from the rest. Moreover, the treatment of the collisional part of the kinetic equation rarely considers inhomogeneous magnetic field. 

To address this ambiguity, and to raise awareness of the current state of the field, our contribution presents a method to model Coulomb collisions consistently in the guiding-center phase-space. Our approach relaxes the issues above, and in the limit of homogeneous magnetic field, the guiding-center formalism presented in this paper reduces to the standard velocity space operators. In addition, the approach directly reveals an aspect that is not always considered~\cite{boozer1981,vernay} though it should: in agreement with~\cite{tessarotto1994}, the stochastic solution directly suggests the collision operators to be implemented in the same phase-space coordinates as the equations of motion. Moreover, it becomes clear that applying collisions consistently with the guiding-center formalism leads to changes not only in the guiding-centers momentum but in its position as well. The significance of the spatial shift will depend on the application, but neglecting it may be hazardous. We also point out that if stochastic methods are applied to solve the kinetic equation, it is not consistent to approximate the collisional part with uniform magnetic field, and simultaneously include magnetic drifts in the equations of motion: part of the deterministic guiding-center motion results from the guiding-center Coulomb drag where the same magnetic field inhomogeneity is present.

 We begin the paper by giving a short introduction on how the simulation of minority species can be done in a particle phase-space, and briefly summarize previous work leading to the guiding-center Fokker-Planck operator~\cite{brizard2004}, which this contribution extends with Monte Carlo methods. The rest of the paper then focuses on constructing stochastic differential equations suitable for solving the guiding-center transformed kinetic equation.

\section{Review of Fokker-Planck Formalism}

In the non-canonical particle phase-space, $\bm{z}=(\bm{x},\bm{p} = m\bm{v})$, the kinetic equation describing the evolution of the distribution function due to Coulomb collisions is often taken to be
\begin{align}
\label{eq:kinetic_equation}
\frac{df}{dt}(\bm{z},t) \;\equiv\; \frac{\partial f}{\partial t} \;+\; \frac{dz^{\alpha}}{dt}\;\frac{\partial f}{\partial z^{\alpha}} = -\;
\frac{\partial}{\partial p^i}\left(K^{i}f-D^{ij}\frac{\partial f}{\partial p^j}\right),
\end{align}
where the collisional momentum friction, or drag, $K^i=\sum_{b}K^i_{ab}$, and the collisional momentum diffusion $D^{ij}=\sum_{b}D^{ij}_{ab}$ are defined with Rosenbluth potentials $(H,G)$ \cite{rosenbluth1957}:
\begin{align}
\label{eq:particle_coulomb_drag}
K^i_{ab}=&\frac{\partial}{\partial p^i}\left(\Gamma_{ab}\frac{m_a}{m_b}\int d^3p'\frac{f_b(\bm{p}')}{\lvert \bm{v}-\bm{v}'\rvert }\right)=\frac{\partial H_{ab}}{\partial p^i}\\
\label{eq:particle_coulomb_diffusion}
D^{ij}_{ab}=&\frac{1}{2}\frac{\partial^2}{\partial p^i\partial p^j}\left(m_a^2\Gamma_{ab}\int d^3p'f_b(\bm{p}')\lvert\bm{v}-\bm{v}'\rvert\right)=\frac{1}{2}\frac{\partial^2G_{ab}}{\partial p^i\partial p^j}.
\end{align}
Here $\Gamma_{ab}=e_a^2e_b^2\ln\Lambda/\epsilon_o$, and $a$ and $b$ refer to test particle and field particles, respectively. The diffusion and friction coefficients satisfy the relation
\begin{equation}
\frac{\partial}{\partial p^{i}}\;D^{ij}_{ab} \;=\; \frac{m_{b}}{m_{a}}\;K^{j}_{ab},
\label{eq:div_D}
\end{equation}
which is useful in proving the momentum and energy conservation properties of the Fokker-Planck collision operator. In the case of isotropic field particle distributions, the Rosenbluth potentials $H_{ab}(\bm{z})=H_{ab}(\bm{x},p)$ and $G_{ab}(\bm{z})=G_{ab}(\bm{x},p)$ become functions of particle position and energy only, and the friction and diffusion coefficient can be written
\begin{align}
\label{eq:prtK}
\bm{K}_{ab}=&\nu_{ab}\;\bm{p},\\
\label{eq:prtD}
\bm{D}_{ab}=&D_{\parallel,ab}\frac{\bm{p}\bm{p}}{p^2}+D_{\perp,ab}(\bm{I}-\frac{\bm{p}\bm{p}}{p^2}),
\end{align}
where $\nu_{ab}=-(1/p)H_{ab}'(p)$, $D_{\parallel,ab}=(1/2)G_{ab}''(p)$ (directed along $\bm{p}$), and $D_{\perp,ab}=(1/2p)G_{ab}'(p)$ (directed perpendicular to $\bm{p}$). 

In order to solve the test particle kinetic equation~(\ref{eq:kinetic_equation}), we consider a general partial differential equation of the form
\begin{align}
\label{eq:general_fp}
\frac{\partial f}{\partial t}(\bm{z},t)=-\frac{\partial}{\partial\bm{z}}\cdot\left[\bm{a}(\bm{z},t)f(\bm{z},t)\right]+\frac{\partial}{\partial\bm{z}}\frac{\partial}{\partial\bm{z}}:\left[\bm{c}(\bm{z},t)f\right],
\end{align}
also known as the Kolmogorov forward equation or Fokker-Planck equation. It describes the temporal evolution of the probability density for finding an individual particle at some phase-space location $\bm{z}$, when the motion of individual particle is governed by the stochastic differential equation
\begin{align}
\label{eq:mc}
dz^{\alpha}=a^{\alpha}(\bm{z},t)dt+\sigma^{\alpha\beta}(\bm{z},t)d\mathcal{W}^{\beta},
\end{align}
where the matrix $\sigma^{\alpha\beta}$ satisfies
\begin{align}
c^{\alpha\beta}=\frac{1}{2}\sigma^{\alpha\gamma}\sigma^{\beta\gamma},
\end{align}
and $\mathcal{W}^{\alpha}$ are independent standard Wiener processes with zero mean and variance $t$. This link between stochastic processes and partial differential equations has been recognized since the work of Kolmogorov~\cite{kolmogorov1931,kolmogorov1933}.

Rearranging the kinetic equation into Kolmogorov forward form (\ref{eq:general_fp}):
\begin{align}
\frac{\partial f}{\partial t} \equiv -\frac{\partial}{\partial\bm{x}}\cdot\left(\bm{v}f\right)-\frac{\partial}{\partial\bm{p}}\cdot\left[ e\left(\bm{E}+\frac{\bm{p}}{m}\times\bm{B}\right)f+\left(\bm{K}+\frac{\partial}{\partial\bm{p}}\cdot\bm{D}\right)f\right]+\frac{\partial}{\partial\bm{p}}\frac{\partial}{\partial\bm{p}}:\left(\bm{D}f\right),
\end{align}
the connection to stochastic motion becomes evident. Applying Eqs.~(\ref{eq:div_D})-(\ref{eq:prtD}), the stochastic differential equations for test particle position and momentum can be written as
\begin{align}
\label{eq:dxprt}
d\bm{x}&=\bm{v}dt\\
\label{eq:dpprt}
d\bm{p}&=\left[e\left(\bm{E}+\frac{\bm{p}}{m}\times\bm{B}\right)+\tilde{\nu}\,\bm{p}\right] dt+\left[\sqrt{2D_{\parallel}}\frac{\bm{p}\bm{p}}{p^2}+\sqrt{2D_{\perp}}\left(\bm{I}-\frac{\bm{p}\bm{p}}{p^2}\right)\right]\cdot d\bm{W},
\end{align}
where $\tilde{\nu}=\sum_{b}\nu_{ab}(1+m_b/m_a)$.

The solution to Eq.~(\ref{eq:kinetic_equation}) can, thus, be found simulating random paths of $z^{\alpha}$ in phase-space according to the dynamics given by Eqs.~(\ref{eq:dxprt}) and~(\ref{eq:dpprt}), and constructing the probability density from the simulated paths.

\section{Guiding-center Fokker-Planck operator}
\label{sec:fokkerplanck}

In 2004, Brizard used Lie-transform methods to first eliminate fast time scales from a general bilinear collision operator, and then applied the method to obtain a guiding-center Fokker-Planck collision operator for a reduced guiding-center phase-space~\cite{brizard2004}, i.e., for a phase-space lacking the gyroangle $\theta$. The resulting operator is expressed in a phase-space divergence form and acts on the gyroangle averaged guiding-center distribution function $\mathcal{F}$ as
\begin{align}
\label{eq:gcFP}
\mathcal{C}_{gcFP}[\mathcal{F}]=-\frac{1}{\mathcal{J}}\frac{\partial}{\partial \mathcal{Z}^{\alpha}}\left[\mathcal{J}\left(\mathcal{K}_{gc}^{\alpha}\mathcal{F}-\mathcal{D}_{gc}^{\alpha\beta}\frac{\partial \mathcal{F}}{\partial \mathcal{Z}^{\beta}}\right)\right],
\end{align}
where $\mathcal{J}$ is the guiding-center phase-space Jacobian. The reduced phase-space friction and diffusion coefficients, $\mathcal{K}^{\alpha}_{gc}$ and $\mathcal{D}^{\alpha\beta}_{gc}$, are $\theta$-averaged projections of the transformed particle momentum-space Fokker-Planck coefficients to the guiding-center phase space, namely
\begin{align}
\label{eq:gcK}
\mathcal{K}^{\alpha}_{gc}&=\langle\bm{T}^{-1}_{gc}\bm{K}\cdot\bm{\Delta}^{\alpha}\rangle\\
\label{eq:gcD}
\mathcal{D}^{\alpha\beta}_{gc}&=\langle(\bm{\Delta}^{\alpha})^{\dagger}\cdot\bm{T}^{-1}_{gc}\bm{D}\cdot\bm{\Delta}^{\beta}\rangle.
\end{align}
The projection vectors involved are defined in terms of the guiding-center Poisson-bracket
\begin{align}
\bm{\Delta}^{\alpha}=\{\bm{T}^{-1}_{gc}\bm{x},\mathcal{Z}^{\alpha}\}_{gc},
\end{align}
and the guiding-center transformations are expressed with the guiding-center push-forward operator, $\bm{T}^{-1}_{gc}$.

\section{Guiding-center friction and diffusion coefficients}
For a phase-space consisting of spatial location $\bm{X}$, energy $\mathcal{E}$, and magnetic moment $\mu$, the gyro-angle averaged guiding-center collisional (isotropic) diffusion coefficients are~\cite{brizard2004}
\begin{align}
\label{eq:Dxx}
\mathcal{D}_{gc}^{\bm{X}\bm{X}}&=\left[(D_{\parallel}-D_{\perp})\frac{\mu B}{2\mathcal{E}}+D_{\perp}\right]\frac{\bm{I}-\bm{\hat{b}}\bm{\hat{b}}}{(m\Omega_{\parallel}^{\star})^2},\\
\label{eq:Dee}
\mathcal{D}_{gc}^{\mathcal{E}\mathcal{E}}&=\frac{2\mathcal{E}}{m}D_{\parallel},\\
\label{eq:Duu}
\mathcal{D}_{gc}^{\mu\mu}&=(1-\epsilon\lambda)\frac{2\mu}{mB}\left[(D_{\parallel}-D_{\perp})\frac{\mu B}{\mathcal{E}}+D_{\perp}\right],\\
\label{eq:Dxe}
\mathcal{D}_{gc}^{\bm{X}\mathcal{E}}&=-\frac{D_{\parallel}}{m}\frac{\bm{\hat{b}}}{\Omega_{\parallel}^{\star}}\times\bm{v}_{gc},\\
\label{eq:Dxu}
\mathcal{D}_{gc}^{\bm{X}\mu}&=-(D_{\parallel}-D_{\perp})\frac{\mu}{2m\mathcal{E}}\frac{\bm{\hat{b}}}{\Omega_{\parallel}^{\star}}\times\bm{v}_{gc},\\
\label{eq:Deu}
\mathcal{D}_{gc}^{\mathcal{E}\mu}&=(2-\epsilon\lambda)D_{\parallel}\frac{\mu}{m},
\end{align}
and the collisional (isotropic) friction coefficients become
\begin{align}
\label{eq:Kx}
\mathcal{K}_{gc}^{\bm{X}}&=\nu\frac{\bm{\hat{b}}}{\Omega_{\parallel}^{\star}}\times\bm{v}_{gc},\\
\label{eq:Ke}
\mathcal{K}_{gc}^{\mathcal{E}}&=-2\nu\mathcal{E},\\
\label{eq:Ku}
\mathcal{K}_{gc}^{\mu}&=-(2-\epsilon\lambda)\nu\mu,
\end{align}
where in Eqs.~(\ref{eq:Duu}),~(\ref{eq:Deu}) and~(\ref{eq:Ku}), $\lambda=(v_{\parallel}/\Omega)\bm{\hat{b}}\cdot\nabla\times\bm{\hat{b}}$ is the guiding-center vorticity parameter. The guiding-center phase-space Jacobian,
\begin{align}
\label{eq:Jeu}
\mathcal{J}_{\mathcal{E}\mu}=mB_{\parallel}^{\star}/\lvert v_{\parallel}\rvert,
\end{align}
is defined in terms of the functions $B_{\parallel}^{\star}(\bm{X},\mathcal{E},\mu)=B+\epsilon (mv_{\parallel}/e)\bm{\hat{b}}\cdot\nabla\times\bm{\hat{b}}$ and $\lvert v_{\parallel}\rvert=\sqrt{(2/m)(\mathcal{E}-\mu B)}$, and the ordering parameter $\epsilon$ is used to clarify the order of each term. The modified gyrofrequency is $\Omega_{\parallel}^{\star}=eB_{\parallel}^{\star}/m$, and the guiding-center velocity is 
\begin{align}
\bm{v}_{gc}=v_{\parallel}\bm{\hat{b}}+\frac{\epsilon\bm{\hat{b}}}{m\Omega_{\parallel}^{\star}}\times\left(\mu\nabla B+mv_{\parallel}^2\bm{\hat{b}}\cdot\nabla\bm{\hat{b}}\right).
\end{align}
It should be noted that the terms~(\ref{eq:Dxe}), (\ref{eq:Dxu}), and~(\ref{eq:Kx}) are of first order because the perpendicular guiding-center velocity 
$(\bm{\hat{b}}\times\bm{v}_{gc}$) is of first order.

Numerical schemes using the pair ($\mathcal{E},\mu$) may, however, experience difficulties as Eq.~(\ref{eq:Jeu}) has a singularity at $\mathcal{E}=\mu B$, i.e., at the turning point of a banana orbit. Instead of the pair ($\mathcal{E},\mu$), we select, for now, the momentum $p$ and the pitch variable 
$\zeta=v_{\parallel}/v$, which are functions of the guiding-center energy and magnetic moment:
\begin{align}
\mathcal{E}&=\frac{p^2}{2m},\quad \frac{\mu B}{\mathcal{E}}=1-\zeta^2.
\end{align}
The momentum $p$ is a convenient choice also because the Rosenbluth potentials are functions of it (for an isotropic field-particle distribution). Reconstructing the guiding-center Lagrange matrix, and calculating the square root of the determinant, our choice gives a new phase-space Jacobian
\begin{align}
\mathcal{J}_{p\zeta}\equiv\sqrt{\det{\omega_{ij}}}=p^2\frac{B_{\parallel}^{\star}}{B},
\end{align}
which is well behaving for all ($\bm{X},p,\zeta$), and we can write $v_{\parallel}=\zeta p/m$.

To calculate the guiding-center friction and diffusion coefficients for the new phase-space, we note the chain rule for a Poisson bracket, $\{F,\mathcal{Z}^{\beta}\}=\{F,\mathcal{Z}^{\alpha}\}\frac{\partial \mathcal{Z}^{\beta}}{\partial \mathcal{Z}^{\alpha}}$, allowing us to write new projection vectors in terms of the old ones as $\bm{\Delta}^{\beta}=\bm{\Delta}^{\alpha}\frac{\partial \mathcal{Z}^{\beta}}{\partial \mathcal{Z}^{\alpha}}$, and to obtain 
\begin{align}
\mathcal{K}_{gc}^{\alpha}&=\mathcal{K}_{gc}^{\gamma}\frac{\partial \mathcal{Z}^{\alpha}}{\partial \mathcal{Z}^{\gamma}},\\
\mathcal{D}_{gc}^{\alpha\beta}&=\frac{\partial \mathcal{Z}^{\alpha}}{\partial \mathcal{Z}^{\gamma}}\mathcal{D}_{gc}^{\gamma\delta}\frac{\partial \mathcal{Z}^{\beta}}{\partial \mathcal{Z}^{\delta}}.
\end{align}
Then, it is a simple task to calculate the partial derivatives between phase-spaces $\mathcal{Z}^{\alpha}=(\bm{X},p,\zeta)$ and $\mathcal{Z}^{\gamma}=(\bm{X},\mathcal{E},\mu)$, and to obtain, from Eqs.~(\ref{eq:Dee})-(\ref{eq:Deu}), the new collisional diffusion coefficients
\begin{align}
\label{eq:Dpp}
\mathcal{D}_{gc}^{pp}&=D_{\parallel},\\
\label{eq:Dzetazeta}
\mathcal{D}_{gc}^{\zeta\zeta}&=(1-\zeta^2)\frac{D_{\perp}}{p^2}(1-\epsilon\lambda),\\
\label{eq:Dxp}
\mathcal{D}_{gc}^{\bm{X}p}&=-\frac{D_{\parallel}}{p}\frac{\bm{\hat{b}}}{\Omega_{\parallel}^{\star}}\times\bm{v}_{gc},\\
\label{eq:Dxzeta}
\mathcal{D}_{gc}^{\bm{X}\zeta}&=\epsilon\frac{\zeta (1-\zeta^2)}{2(m\Omega_{\parallel}^{\star})^2}\left[(D_{\parallel}+D_{\perp})\bm{\hat{b}}\cdot\nabla\bm{\hat{b}}-D_{\perp}\nabla_{\perp}\ln B\right],\\
\label{eq:Dzetap}
\mathcal{D}_{gc}^{\zeta p}&=\frac{1-\zeta^2}{2\zeta}\frac{D_{\parallel}}{p}\epsilon\lambda,
\end{align}
where $\nabla_{\perp}=(\bm{I}-\bm{\hat{b}}\bm{\hat{b}})\cdot\nabla$, as well as, from Eqs.~(\ref{eq:Ke})-(\ref{eq:Ku}) the new friction coefficients
\begin{align}
\label{eq:Kp}
\mathcal{K}_{gc}^{p}&=-\nu p,\\
\label{eq:Kzeta}
\mathcal{K}_{gc}^{\zeta}&=-\left(\frac{1-\zeta^2}{2\zeta}\right)\;\nu\epsilon\lambda.
\end{align}
The coefficients (\ref{eq:Dxx}) and (\ref{eq:Kx}) for $\mathcal{D}_{gc}^{\bm{X}\bm{X}}$ and $\mathcal{K}_{gc}^{\bm{X}}$ remain unchanged.

We will use the $(p,\zeta)$ phase-space for comparison with the standard model~\cite{boozer1981}, but to consider the whole kinetic equation instead of only the right-hand-side, we would rather use the phase-space $\mathcal{Z}^{\alpha}=(\bm{X},v_{\parallel},\mu)$. This is a convenient choice because the equations of motion are often expressed for this particular set of coordinates. Again, transforming the friction and diffusion coefficients as previously, we obtain a new set of coefficients
\begin{align}
\label{eq:Dvv}
\mathcal{D}_{gc}^{v_{\parallel}v_{\parallel}}&=\frac{D_{\parallel}}{m^2}+(1-\epsilon\lambda)\frac{D_{\perp}-D_{\parallel}}{m^2}\frac{\mu B}{\mathcal{E}},\\
\label{eq:Dxv}
\mathcal{D}_{gc}^{\bm{X}v_{\parallel}}&=\frac{\epsilon v_{\parallel}}{(m\Omega_{\parallel}^{\star})^2}(D_{\parallel}-D_{\perp})\frac{\mu B}{2\mathcal{E}}\nabla_{\perp}\ln{B}\nonumber\\&+\frac{\epsilon v_{\parallel}}{(m\Omega_{\parallel}^{\star})^2}\left[D_{\parallel}+\frac{\mu B}{2\mathcal{E}}(D_{\perp}-D_{\parallel})\right]\bm{\hat{b}}\cdot\nabla\bm{\hat{b}},\\
\label{eq:Dvu}
\mathcal{D}_{gc}^{\mu v_{\parallel}}&=(1-\epsilon\lambda)\frac{\mu v_{\parallel}}{m\mathcal{E}}(D_{\parallel}-D_{\perp})+\epsilon\lambda\frac{\mu}{v_{\parallel}m^2}D_{\parallel},\\
\label{eq:Kv}
\mathcal{K}_{gc}^{v_{\parallel}}&=-\nu v_{\parallel}-\epsilon\lambda\frac{\mu B}{mv_{\parallel}}\nu,
\end{align}
with a phase-space Jacobian $\mathcal{J}_{v_{\parallel}\mu}=m^2B_{\parallel}^{\star}$.

\section{Monte Carlo method for collisional guiding-center Kinetic equation}
The gyroaveraged guiding-center kinetic equation now stands
\begin{align}
\label{eq:gcKinetic}
\frac{\partial\mathcal{F}}{\partial t}+\dot{\mathcal{Z}}^{\alpha}\frac{\partial\mathcal{\mathcal{F}}}{\partial \mathcal{Z}^{\alpha}}=-\frac{1}{\mathcal{J}}\frac{\partial}{\partial \mathcal{Z}^{\alpha}}\left[\mathcal{J}\left(\mathcal{K}^{\alpha}\mathcal{F}-\mathcal{D}^{\alpha\beta}\frac{\partial \mathcal{F}}{\partial \mathcal{Z}^{\beta}}\right)\right],
\end{align}
where $\dot{\mathcal{Z}}^{\alpha}$ is the equation of motion for the phase-space coordinate $\mathcal{Z}^{\alpha}$. The resemblance to the test particle kinetic equation~(\ref{eq:kinetic_equation}) is obvious and, thus, the method for obtaining the solution should be similar as well. If we apply the Liouville theorem  
\begin{align}
\frac{\partial\mathcal{J}}{\partial t}+\frac{\partial}{\partial \mathcal{Z}^{\alpha}}\left(\mathcal{J}\dot{\mathcal{Z}}^{\alpha}\right)=0,
\end{align}
 and restrict ourselves into time-independent guiding-center transformation, i.e., $\partial\mathcal{J}/\partial t=0$, also the guiding-center kinetic equation can be written in the form of Kolmogorov forward equation as 
\begin{align}
\label{eq:gcKolmogorovForward}
\frac{\partial \mathcal{F}}{\partial t}=&-\frac{1}{\mathcal{J}}\frac{\partial}{\partial \mathcal{Z}^{\alpha}}\left(\mathcal{J}\mathcal{A}^{\alpha}\mathcal{F}\right)+\frac{1}{\mathcal{J}}\frac{\partial^2}{\partial \mathcal{Z}^{\alpha}\partial \mathcal{Z}^{\beta}}\left(\mathcal{J}\mathcal{D}^{\alpha\beta}\mathcal{F}\right),
\end{align}
where the coefficient $\mathcal{A}^{\alpha}$ is 
\begin{align}
\label{eq:gcdrift}
\mathcal{A}^{\alpha}=\dot{\mathcal{Z}}^{\alpha}+\mathcal{K}^{\alpha}+\frac{1}{\mathcal{J}}\frac{\partial}{\partial \mathcal{Z}^{\beta}}(\mathcal{J}\mathcal{D}^{\alpha\beta}).
\end{align}
The stochastic differential equation for a phase-space coordinate $\mathcal{Z}^{\alpha}$ thus becomes
\begin{align}
\label{eq:gcmc}
d\mathcal{Z}^{\alpha}=\mathcal{A}^{\alpha}dt+\Sigma^{\alpha\beta}d\mathcal{W}^{\beta},
\end{align}
where the matrix $\Sigma^{\alpha\beta}$ satisfies
\begin{align}
\mathcal{D}^{\alpha\beta}=\frac{1}{2}\Sigma^{\alpha\gamma}\Sigma^{\beta\gamma},
\end{align}
and 
the solution to the guiding-center kinetic equation can be found similarly as in the particle phase-space: simulating random paths of $\mathcal{Z}^{\alpha}$ according to the dynamics given by Eq.~(\ref{eq:gcmc}), and constructing the probability density from the simulated paths.

In particle phase-space it was straight-forward to obtain the matrix $\sigma^{\alpha\beta}$, but in guiding-center phase-space the decomposition of $\mathcal{D}^{\alpha\beta}$ is not trivial. $\Sigma^{\alpha\beta}$ could be obtained with eigenvalue decomposition, if all the entries of $\mathcal{D}^{\alpha\beta}$ had equal units and $\mathcal{D}^{\alpha\beta}$ was positive definite. Unfortunately, the units differ as the units of $\mathcal{D}^{\alpha\beta}$ are the unit of $\mathcal{Z}^{\alpha}$ times the unit of $\mathcal{Z}^{\beta}$ divided by second, and in the case of arbitrary magnetic field it is difficult to prove that $\mathcal{D}^{\alpha\beta}$ is positive definite for all $\mathcal{Z}^{\alpha}$. We can, however, consider the problem in terms of the guiding-center ordering parameter~$\epsilon$.

\subsection{Zeroth-order method for $(p,\zeta)$}
If we restrict ourselves to zeroth order in magnetic field non-uniformity, and use the phase-space $\mathcal{Z}^{\alpha}=(\bm{X},p,\zeta)$, the diffusion tensor forms a block-diagonal 
\begin{align}
\bm{\mathcal{D}}^{\alpha\beta} =
\left( \begin{array}{ccc}
\mathcal{D}^{\bm{X}\bm{X}} & 0 & 0 \\
0 & D_{\parallel} & 0 \\
0 & 0 & (1-\zeta^2)\frac{D_{\perp}}{p^2}
\end{array} \right),
\end{align}
where the velocity space block is strictly diagonal and separated from the spatial block. As $\mathcal{D}^{\bm{X}\bm{X}}$, given by Eq.~(\ref{eq:Dxx}), is non-negative, the decomposition of $\mathcal{D}^{\alpha\beta}$ is straight-forward, and we may write
\begin{align}
\bm{\Sigma}^{\alpha\beta}=
\left( \begin{array}{ccc}
\Sigma^{\bm{X}\bm{X}} & 0 & 0 \\
0 & \sqrt{2D_{\parallel}} & 0 \\
0 & 0 & \sqrt{(1-\zeta^2)\frac{2D_{\perp}}{p^2}}
\end{array} \right),
\end{align}
where the spatial block is given by
\begin{align}
\Sigma^{\bm{X}\bm{X}}=\sqrt{2D^{\bm{X}}}(\bm{I}-\bm{\hat{b}}\bm{\hat{b}}), 
\end{align}
with a spatial diffusion coefficient
\begin{align}
\label{eq:dx}
D^{\bm{X}}=[(D_{\parallel}-D_{\perp})\frac{\mu B}{2\mathcal{E}}+D_{\perp}]/[(m\Omega_{\parallel}^{\star})^2].
\end{align}

Applying the zeroth order approximation also for the guiding-center friction vector, which simplifies to
\begin{align}
\bm{\mathcal{K}}^{\alpha}=
\left( \begin{array}{c}
\bm{0} \\
-\nu p \\
0
\end{array} \right),
\end{align}
the stochastic differential equations for the phase-space $\mathcal{Z}^{\alpha}=(\bm{X},p,\zeta)$ then become
\begin{align}
\label{eq:zerodx}
d\bm{X}=&\bm{v}_{gc}dt+\sqrt{2D^{\bm{X}}}\left(\bm{I}-\bm{\hat{b}}\bm{\hat{b}}\right)\cdot d\bm{\mathcal{W}}^{\bm{X}},\\
\label{eq:zerodp}
dp=&\left(\dot{p}-\nu p+\frac{\partial D_{\parallel}}{\partial p}+2\frac{D_{\parallel}}{p}\right)dt+\sqrt{2D_{\parallel}}d\mathcal{W}^{p},\\
\label{eq:zerodzeta}
d\zeta=&\left(\dot{\zeta}-\zeta\frac{2D_{\perp}}{p^2}\right)dt+\sqrt{(1-\zeta^2)\frac{2D_{\perp}}{p^2}}d\mathcal{W}^{\zeta}.
\end{align}
We note that the phase-space spatial position $\bm{X}$ travels deterministically with the guiding-center velocity $\bm{v}_{gc}$, as expected, but diffuses randomly in a plane perpendicular to the local magnetic field with a spatial diffusion coefficient $D^{\bm{X}}$. The same diffusion coefficient for the spatial transport is reported also in~\cite{vernay}. Additionally, if we neglect $\dot{\zeta}$, and apply Euler method for solving what remains of Eq.~(\ref{eq:zerodzeta}), we exactly recover the standard Lorentz operator for the pitch scattering. Similarly, constructing an operator for the energy with $d\mathcal{E}=pdp/m$, also the change in the energy becomes equivalent to the standard expression. This observation makes sense because the guiding-center transformation is scalar invariant and, thus, the guiding-center energy should equal the particle energy up to first order~\cite{brizard2004}. 

We note that the guiding-centers tend not to be followed with the coordinates $(p,\zeta)$. Use of pitch and momentum operators without the corresponding equations of motion is inconsistent with the idea of solving the kinetic equation with random processes and, thus, we see fit to consider coordinates $(v_{\parallel},\mu)$ in which the equations of motion most often are implemented.

\subsection{Zeroth-order method for $(v_{\parallel},\mu)$}
In phase-space $\mathcal{Z}^{\alpha}=(\bm{X},v_{\parallel},\mu)$, the zeroth order guiding-center diffusion tensor takes the form
\begin{align}
\bm{\mathcal{D}}^{\alpha\beta} =
\left( \begin{array}{ccc}
\mathcal{D}^{\bm{X}\bm{X}} & 0 & 0 \\
0 & \mathcal{D}^{v_{\parallel}v_{\parallel}} & \mathcal{D}^{v_{\parallel}\mu} \\
0 & \mathcal{D}^{v_{\parallel}\mu} & \mathcal{D}^{\mu\mu}
\end{array} \right),
\end{align}
where the velocity block is no longer diagonal. The spatial and velocity blocks are still disconnected,  
and the motion of $\bm{X}$ is determined by Eq.~(\ref{eq:zerodx}), but the coordinates $v_{\parallel}$ and $\mu$, however, require more attention.

Due to symmetry, $\mathcal{D}^{\alpha\beta}$ is expressible as a product
\begin{align}
\mathcal{D}^{\alpha\beta}=\mathcal{B}^{\alpha\gamma}\mathcal{Y}^{\gamma\nu}\mathcal{B}^{\nu\beta},
\end{align}
where $\mathcal{B}^{\alpha\beta}$ is a diagonal matrix defined so that the entries of $\mathcal{Y}^{\alpha\beta}$ have equal units. Now focusing to the velocity block only, we should choose $\mathcal{B}^{v_{\parallel}v_{\parallel}}$ to have units of $v_{\parallel}$, and $\mathcal{B}^{\mu\mu}$ to have units of $\mu$. An intuitive choice would be ($\mathcal{B}^{v_{\parallel}v_{\parallel}}=v_{\parallel},\mathcal{B}^{\mu\mu}=\mu$) but as both of these coordinates may obtain a value zero for some guiding-center orbit topology, $\mathcal{Y}^{\alpha\beta}$ would be ill-defined at that phase-space position. Thus, instead, we choose 
\begin{align}
\mathcal{B}^{v_{\parallel}v_{\parallel}}=v,\quad \mathcal{B}^{\mu\mu}=\mathcal{E}/B,
\end{align}
which are always positive for non-zero kinetic energy.

Neglecting the first order terms, the velocity components for the normalized symmetric matrix $\mathcal{Y}^{\alpha\beta}$ become
\begin{align}
\mathcal{Y}^{v_{\parallel}v_{\parallel}}&=\frac{1}{m^2v^2}\left[ D_{\parallel}\zeta^2+D_{\perp}(1-\zeta^2)\right]\\
\mathcal{Y}^{v_{\parallel}\mu}&=\frac{2}{m^2v^2}\zeta (1-\zeta^2)(D_{\parallel}-D_{\perp}) \\
\mathcal{Y}^{\mu\mu}&=\frac{4}{m^2v^2}(1-\zeta^2)\left[D_{\parallel}(1-\zeta^2)+D_{\perp}\zeta^2\right],
\end{align}
with eigenvalues
\begin{align}
\label{eq:eigenvalues_zero_order}
\lambda=&\frac{1}{2}\left[\mathcal{Y}^{v_{\parallel}v_{\parallel}}+\mathcal{Y}^{\mu\mu}\pm\sqrt{\left(\mathcal{Y}^{v_{\parallel}v_{\parallel}}+\mathcal{Y}^{\mu\mu}\right)^2-4\left[\mathcal{Y}^{v_{\parallel}v_{\parallel}}\mathcal{Y}^{\mu\mu}-(\mathcal{Y}^{v_{\parallel}\mu})^2\right]}\right]\nonumber\\
=&\frac{1}{2}\left[\mathcal{Y}^{v_{\parallel}v_{\parallel}}+\mathcal{Y}^{\mu\mu}\pm\sqrt{\left(\mathcal{Y}^{v_{\parallel}v_{\parallel}}-\mathcal{Y}^{\mu\mu}\right)^2+4(\mathcal{Y}^{v_{\parallel}\mu})^2}\right],
\end{align}
that are always real, non-negative, and non-degenerate because the determinant satisfies
\begin{align}
\mathcal{Y}^{v_{\parallel}v_{\parallel}}\mathcal{Y}^{\mu\mu}-(\mathcal{Y}^{v_{\parallel}\mu})^2=\frac{4}{m^4v^4}(1-\zeta^2)D_{\parallel}D_{\perp}\ge 0.
\end{align}
The only zero eigenvalue is obtained at $\zeta=\pm 1$, in which case the velocity block of $\mathcal{Y}^{\alpha\beta}$ is already diagonal with the only non-zero entry being $\mathcal{Y}^{v_{\parallel}v_{\parallel}}=D_{\parallel}/(m^2v^2)$.

Considering the case $\zeta=\pm 1$ separately, we may now give the expression for the zeroth order $\Sigma^{\alpha\beta}$ . For $\lvert\zeta\rvert<1$ we have
\begin{align}
\bm{\Sigma}^{\alpha\beta} =
\left( \begin{array}{cc}
\sqrt{2D^{\bm{X}}}(\bm{I}-\bm{\hat{b}}\bm{\hat{b}}) & \bm{0} \\
\bm{0} & \left( \begin{array}{cc}
v & 0 \\
0 & \mathcal{E}/B
\end{array} \right)\left( \begin{array}{cc}
\bm{Q}_1 & \bm{Q}_2 \end{array} \right)
\left( \begin{array}{cc}
\sqrt{2\lambda_1} & 0 \\
0 & \sqrt{2\lambda_2}
\end{array} \right)
\end{array} \right),
\end{align}
where $\lambda_1$ and $\lambda_2$ are the eigenvalues obtained from Eq.~(\ref{eq:eigenvalues_zero_order}), and the vectors $\bm{Q}_1$ and $\bm{Q}_2$ are the normalized eigenvectors related to the corresponding eigenvalues 
\begin{align}
\bm{Q}_{1}&=\frac{1}{\sqrt{1+\left(\frac{\mathcal{Y}^{v_{\parallel}\mu}}{\lambda_1-\mathcal{Y}^{\mu\mu}}\right)^2}}\left( \begin{array}{c}
1\\
\frac{\mathcal{Y}^{v_{\parallel}\mu}}{\lambda_1- \mathcal{Y}^{\mu\mu}}
\end{array} \right),\\
\bm{Q}_{2}&=\frac{1}{\sqrt{1+\left(\frac{\mathcal{Y}^{v_{\parallel}\mu}}{\lambda_2-\mathcal{Y}^{v_{\parallel}v_{\parallel}}}\right)^2}}\left( \begin{array}{c}
\frac{\mathcal{Y}^{v_{\parallel}\mu}}{\lambda_2- \mathcal{Y}^{v_{\parallel}v_{\parallel}}}\\
1
\end{array} \right).
\end{align}
For the limiting case $\zeta=\pm 1$, although this rarely happens in numerical simulations, we obtain 
\begin{align}
\bm{\Sigma}^{\alpha\beta} =
\left( \begin{array}{cc}
\sqrt{2D^{\bm{X}}}(\bm{I}-\bm{\hat{b}}\bm{\hat{b}}) & \bm{0} \\
\bm{0} & \left( \begin{array}{cc}
\frac{\sqrt{2D_{\parallel}}}{m} & 0 \\
0 & 0
\end{array} \right)
\end{array} \right).
\end{align}
In order to complete the zeroth order stochastic differential equations for $(v_{\parallel},\mu)$, also the expressions for $\mathcal{A}^{\alpha}$ are required. With the zeroth-order friction vector
\begin{align}
\bm{\mathcal{K}}^{\alpha}=
\left( \begin{array}{c}
\bm{0} \\
-\nu v_{\parallel} \\
-2\nu\mu
\end{array} \right),
\end{align}
the coefficients are explicitly
\begin{align}
\label{eq:Av}
\mathcal{A}^{v_{\parallel}}=&\ \dot{v}_{\parallel}-\nu v_{\parallel}+\frac{v_{\parallel}}{m\mathcal{E}}\left(D_{\parallel}-D_{\perp}+\frac{p}{2}\frac{\partial D_{\parallel}}{\partial p}\right),\\
\label{eq:Amu}
\mathcal{A}^{\mu}=&-2\nu\mu+\frac{\mu}{m\mathcal{E}}\left[ p\frac{\partial D_{\parallel}}{\partial p}+3\left(D_{\parallel}-D_{\perp}\right)\right]+\frac{2D_{\perp}}{mB},
\end{align}
where, in Eq.~(\ref{eq:Amu}), we have neglected $\dot{\mu}$ since the magnetic moment is a constant of Hamiltonian guiding-center motion. It is also important to notice that if $\zeta=\pm 1$, i.e., $\mu=0$, then the stochastic contribution to the change in $\mu$ vanishes and the deterministic contribution is positive. Thus the non-negativity of the magnetic moment is preserved.

Why to bother first order then? To be exact, the ordering applies also to the equations of motion for the phase-space coordinates. In zeroth order, i.e., in the absence of magnetic gradient drifts, the guiding-center velocity would be reduced simply to motion along the field-line. Then bounce-center dynamics in the zero-orbit-width limit with corresponding collision operator would be more relevant.

\subsection{First-order method}
If we include also the first order corrections in $\mathcal{D}^{\alpha\beta}$, caused by the magnetic field inhomogeneity, the spatial part is no longer disconnected from the velocity part, and we have to normalize the whole $\mathcal{D}^{\alpha\beta}$. A convenient choice is to define
\begin{align}
\mathcal{B}^{\alpha\beta}=\left( \begin{array}{cc}
\lvert\bm{X}\rvert\bm{I} & \bm{0} \\
\bm{0} & \left( \begin{array}{cc}
 v & 0 \\
 0 & \mathcal{E}/B
\end{array} \right)
\end{array} \right),
\end{align}
wich yields the normalized matrix components
%
%
\begin{align}
\mathcal{Y}^{\alpha\beta}=\left( \begin{array}{ccc}
\frac{\mathcal{D}^{\bm{X}\bm{X}}}{\lvert\bm{X}\rvert^2} & 
\frac{\mathcal{D}^{\bm{X}v_{\parallel}}}{\lvert\bm{X}\rvert v} &
\frac{B}{\mathcal{E}}\frac{\mathcal{D}^{\bm{X}\mu}}{\lvert\bm{X}\rvert}
\\
\frac{\mathcal{D}^{v_{\parallel}\bm{X}}}{\lvert\bm{X}\rvert v} &
\frac{\mathcal{D}^{v_{\parallel}v_{\parallel}}}{v^2} &
\frac{B}{\mathcal{E}}\frac{\mathcal{D}^{v_{\parallel}\mu}}{v}
\\
\frac{B}{\mathcal{E}}\frac{\mathcal{D}^{\mu\bm{X}}}{\lvert\bm{X}\rvert} &
\frac{B}{\mathcal{E}}\frac{\mathcal{D}^{v_{\parallel}\mu}}{v} &
\frac{B^2\mathcal{D}^{\mu\mu}}{\mathcal{E}^2}
 \end{array} \right).
\end{align}
It is difficult to show the full $\mathcal{Y}^{\alpha\beta}$ to be positive-definite and, in fact, one of the eigenvalues is zero. This is evident as both $\mathcal{D}^{\bm{X}v_{\parallel}}$ and $\mathcal{D}^{\bm{X}\mu}$ are perpendicular to $\bm{\hat{b}}$, and $\mathcal{D}^{\bm{X}\bm{X}}\cdot\bm{\hat{b}}=\bm{0}$, revealing that
\begin{align}
\mathcal{U}^{\alpha}=\left( \begin{array}{c}
\bm{\hat{b}}\\
0\\
0\end{array} \right),
\end{align}
is an eigenvector of $\mathcal{Y}^{\alpha\beta}$ corresponding to eigenvalue $\lambda=0$. Thus $\mathcal{Y}^{\alpha\beta}$ can be at most positive semi-definite.

Since $\mathcal{Y}^{\alpha\beta}$ is symmetric, its eigenvectors are orthogonal, and, thus, the rest of the eigenvectors take the form
\begin{align}
\mathcal{U}^{\alpha}=\left( \begin{array}{c}
\mathcal{U}^{\bm{X}}\\
\mathcal{U}^{v_{\parallel}}\\
\mathcal{U}^{\mu} \end{array} \right),
\end{align}
where $\mathcal{U}^{\bm{X}}$ is a vector perpendicular to $\bm{\hat{b}}$. The eigenvalue equation $\mathcal{Y}^{\alpha\beta}\mathcal{U}^{\beta}=\lambda\mathcal{U}^{\alpha}$ can be solved formally using the fact that $\mathcal{U}^{\bm{X}}\cdot\bm{\hat{b}}=0$: Solving the spatial part of the eigenvector as
\begin{align}
\mathcal{U}^{\bm{X}}=\frac{\frac{\mathcal{D}^{\bm{X}v_{\parallel}}}{\lvert\bm{X}\rvert v}\mathcal{U}^{v_{\parallel}}+\frac{B}{\mathcal{E}}\frac{\mathcal{D}^{\bm{X}\mu}}{\lvert\bm{X}\rvert}\mathcal{U}^{\mu}}{\lambda-\frac{D^{\bm{X}}}{\lvert\bm{X}\rvert^2}},
\end{align}
and eliminating it from the rest of the equations, leads to a $2\times 2$ matrix equation for the velocity components
\begin{align}
\left( \begin{array}{cc}
\frac{\frac{\mathcal{D}^{\bm{X}v_{\parallel}}}{\lvert\bm{X}\rvert v}\cdot\frac{\mathcal{D}^{\bm{X}v_{\parallel}}}{\lvert\bm{X}\rvert v}}{\lambda-\frac{D^{\bm{X}}}{\lvert\bm{X}\rvert^2}}+\frac{\mathcal{D}^{v_{\parallel}v_{\parallel}}}{v^2}
& 
\frac{\frac{\mathcal{D}^{\bm{X}v_{\parallel}}}{\lvert\bm{X}\rvert v}\cdot\frac{B}{\mathcal{E}}\frac{\mathcal{D}^{\bm{X}\mu}}{\lvert\bm{X}\rvert}}{\lambda-\frac{D^{\bm{X}}}{\lvert\bm{X}\rvert^2}}+\frac{B}{\mathcal{E}}\frac{\mathcal{D}^{v_{\parallel}\mu}}{v}\\
\frac{\frac{\mathcal{D}^{\bm{X}v_{\parallel}}}{\lvert\bm{X}\rvert v}\cdot\frac{B}{\mathcal{E}}\frac{\mathcal{D}^{\bm{X}\mu}}{\lvert\bm{X}\rvert}}{\lambda-\frac{D^{\bm{X}}}{\lvert\bm{X}\rvert^2}}+\frac{B}{\mathcal{E}}\frac{\mathcal{D}^{v_{\parallel}\mu}}{v}
&
\frac{\frac{B}{\mathcal{E}}\frac{\mathcal{D}^{\bm{X}\mu}}{\lvert\bm{X}\rvert}\cdot\frac{B}{\mathcal{E}}\frac{\mathcal{D}^{\bm{X}\mu}}{\lvert\bm{X}\rvert}}{\lambda-\frac{D^{\bm{X}}}{\lvert\bm{X}\rvert^2}}+\frac{B^2\mathcal{D}^{\mu\mu}}{\mathcal{E}^2}
\end{array} \right)
\left( \begin{array}{c}
\mathcal{U}^{v_{\parallel}}\\
\mathcal{U}^{\mu}
\end{array}\right)
=\lambda\left( \begin{array}{c}
\mathcal{U}^{v_{\parallel}}\\
\mathcal{U}^{\mu}
\end{array}\right),
\end{align}
which can be solved after obtaining the eigenvalues from a quartic equation
\begin{align}
\label{eq:quartic}
a_4\lambda^4+a_3\lambda^3+a_2\lambda^2+a_1\lambda+a_0=0,
\end{align}
where the coefficients are 
\begin{align}
\label{eq:a4}
a_4=&\
1\\
\label{eq:a3}
a_3=&
-\left( \frac{\mathcal{D}^{v_{\parallel}v_{\parallel}}}{v^2}+\frac{B^2\mathcal{D}^{\mu\mu}}{\mathcal{E}^2}+2\frac{D^{\bm{X}}}{\lvert\bm{X}\rvert^2}\right)
\\
\label{eq:a2}
a_2=&\
2\left(\frac{\mathcal{D}^{v_{\parallel}v_{\parallel}}}{v^2}+\frac{B^2\mathcal{D}^{\mu\mu}}{\mathcal{E}^2}\right)\frac{D^{\bm{X}}}{\lvert\bm{X}\rvert^2}
-
\left(\frac{\mathcal{D}^{\bm{X}v_{\parallel}}}{\lvert\bm{X}\rvert v}\right)^2
-
\left(\frac{B}{\mathcal{E}}\frac{\mathcal{D}^{\bm{X}\mu}}{\lvert\bm{X}\rvert}\right)^2
\nonumber\\
&\
+
\frac{\mathcal{D}^{v_{\parallel}v_{\parallel}}}{v^2}\frac{B^2\mathcal{D}^{\mu\mu}}{\mathcal{E}^2}
+
\left(\frac{D^{\bm{X}}}{\lvert\bm{X}\rvert^2}\right)^2
-
\left(\frac{B}{\mathcal{E}}\frac{\mathcal{D}^{v_{\parallel}\mu}}{v}\right)^2
\\
\label{eq:a1}
a_1=&
\left(\frac{\mathcal{D}^{\bm{X}v_{\parallel}}}{\lvert\bm{X}\rvert v}\right)^2\left(\frac{D^{\bm{X}}}{\lvert\bm{X}\rvert^2}+\frac{B^2\mathcal{D}^{\mu\mu}}{\mathcal{E}^2}\right)
+
\left(\frac{B}{\mathcal{E}}\frac{\mathcal{D}^{\bm{X}\mu}}{\lvert\bm{X}\rvert}\right)^2\left(\frac{D^{\bm{X}}}{\lvert\bm{X}\rvert^2}+\frac{\mathcal{D}^{v_{\parallel}v_{\parallel}}}{v^2}\right)\nonumber\\
&\ -
\left(\frac{D^{\bm{X}}}{\lvert\bm{X}\rvert^2}\right)^2\left(\frac{\mathcal{D}^{v_{\parallel}v_{\parallel}}}{v^2}+\frac{B^2\mathcal{D}^{\mu\mu}}{\mathcal{E}^2}\right)
-
2\frac{B}{\mathcal{E}}\frac{\mathcal{D}^{\bm{X}\mu}}{\lvert\bm{X}\rvert}\cdot\frac{\mathcal{D}^{\bm{X}v_{\parallel}}}{\lvert\bm{X}\rvert v}\frac{B}{\mathcal{E}}\frac{\mathcal{D}^{v_{\parallel}\mu}}{v}\nonumber\\
&\ +2\frac{D^{\bm{X}}}{\lvert\bm{X}\rvert^2}\left( \left(\frac{B}{\mathcal{E}}\frac{\mathcal{D}^{v_{\parallel}\mu}}{v}\right)^2-\frac{\mathcal{D}^{v_{\parallel}v_{\parallel}}}{v^2}\frac{B^2\mathcal{D}^{\mu\mu}}{\mathcal{E}^2}\right)
\\
\label{eq:a0}
a_0=&
\left(\frac{\mathcal{D}^{\bm{X}v_{\parallel}}}{\lvert\bm{X}\rvert v}\right)^2\left(\frac{B}{\mathcal{E}}\frac{\mathcal{D}^{\bm{X}\mu}}{\lvert\bm{X}\rvert}\right)^2
+
\left(\frac{D^{\bm{X}}}{\lvert\bm{X}\rvert^2}\right)^2\frac{B^2\mathcal{D}^{\mu\mu}}{\mathcal{E}^2}\frac{\mathcal{D}^{v_{\parallel}v_{\parallel}}}{v^2}\nonumber\\
&\ -
\frac{D^{\bm{X}}}{\lvert\bm{X}\rvert^2}\left(\left(\frac{\mathcal{D}^{\bm{X}v_{\parallel}}}{\lvert\bm{X}\rvert v}\right)^2\frac{B^2\mathcal{D}^{\mu\mu}}{\mathcal{E}^2}+\left(\frac{B}{\mathcal{E}}\frac{\mathcal{D}^{\bm{X}\mu}}{\lvert\bm{X}\rvert}\right)^2\frac{\mathcal{D}^{v_{\parallel}v_{\parallel}}}{v^2}\right).
\end{align}

According to Descartes' rule of signs, if the terms of a single-variable polynomial with real coefficients are ordered by descending variable exponent, then the number of positive roots of the polynomial is either equal to the number of sign differences between consecutive nonzero coefficients, or is less than it by a multiple of 2. Similarly, the number of negative roots is the number of sign changes after multiplying the coefficients of odd-power terms by −1, or fewer than it by a multiple of 2. For our quartic polynomial, Eq.~(\ref{eq:quartic}), to possibly have four positive roots the rule means that the coefficients must satisfy
\begin{align}
a_4>0,\quad\ a_3<0,\quad\ a_2>0,\quad\ a_1<0,\quad\ a_0>0.
\end{align}
If this is true, then the rule states that the number of negative roots is at most zero, and we have exactly four positive roots. In this case the matrix $\mathcal{Y}^{\alpha\beta}$ will be positive semi-definite, and we will be able to construct the matrix $\Sigma^{\alpha\beta}$ with the aid of the eigenvalues and eigenvectors. Analytical verification of this condition is a difficult task but, if we neglect the second-order terms in the coefficients~(\ref{eq:a4}--\ref{eq:a0}), and assume a toroidal magnetic field $\bm{B}= B_{\phi}\,\nabla\phi$, the condition is satisfied. For an arbitrary magnetic field, however, the condition must be verified numerically.

\section{Conclusions}
We have discovered that in many cases the implementations of a collision operator into a guiding-center following code has led to a situation where, inadvertedly, the particle and guiding-center formalisms are mixed up. The consequences become obvious when the problem is properly looked at as solving the kinetic equation with stochastic approach. The implementations for solving the test particle kinetic equation tend to use different phase-space coordinates for the equations of motion and for the collisional part of the kinetic equation, though stochastic methods clearly suggest that both sides of the kinetic equation need to be treated with the same phase-space. This ambiguity is mainly a heritage from applying the guiding-center transformation only to equations of motion and not to the collisional term, and it also results in the loss of spatial diffusion in the guiding-center motion.

 Solving the kinetic equation with stochastic differential equations also points out that the deterministic motion of the phase-space coordinates consists both of the equations of motion and of the collisional drag. Thus, both these terms should be treated consistently with respect to both the magnetic field inhomogeneity and the choice of numerical integration method. Current implementations, however, tend to assume uniform magnetic field for the collisional contribution and yet include drifts in the equations of motion. Additionally, the equations of motion are integrated with accurate adaptive Runge-Kutta methods whereas the collisional drag is treated with the very crude Euler method.

This contribution relaxes these issues. The guiding-center kinetic equation, the starting point for the present work, is obtained applying the guiding-center Lie-transformation consistently on both sides of the particle kinetic equation. Thus, both sides of the resulting guiding-center kinetic equation are treated consistently with same phase-space coordinates. Also, as the collisional part of the kinetic equation is considered up to the same order in magnetic field non-uniformity as the equations of motion, the guiding-center Coulomb drag appearing in the stochastic differential equations can be treated consistently with the contribution from the equations of motion.

 It was also shown that the standard approach can be recovered if the approximation of uniform field is applied. In addition, the same spatial diffusion coefficient was obtained as reported previously. Considering the inhomogeneous magnetic field, however, it becomes difficult to give analytical expressions for the coefficients needed for the stochastic differential equations. We provide one eigenvalue with the corresponding eigenvector and give a quartic equation for solving the rest of the eigenvalues and eigenvectors. Numerical calculation of the eigen-decomposition for a $5\times 5$ matrix, however, should not be an issue.

\begin{acknowledgments}
This work, supported by the European Communities under the contract of Association between Euratom--Tekes, was carried out within the framework of the European Fusion Development Agreement. The views and opinions expressed herein do not necessarily reflect those of the European Commission. Funding was received also from the Academy of Finland project No. 259675, from the Fusion For Energy Grant 379, from the Emil Aaltonen Foundation, and from the Finnish Foundation for Technology Promotion, making this collaboration possible.
\end{acknowledgments}

\providecommand{\noopsort}[1]{}\providecommand{\singleletter}[1]{#1}%


\end{document}